\documentclass[useAMS,usenatbib,onecolumn]{mn2e}

\usepackage{psfig}
\usepackage{colordvi}
\usepackage{ulem}

\newcommand{\myemail}{agolATastro.washington.edu}

\begin{document}

\title[HD 209458b transits]{A limit on the presence of Earth-mass planets around
a Sun-like star}

\author[Agol \& Steffen]{Eric Agol$^1$\thanks{\myemail}, Jason H. Steffen$^2$,\\
$^1$Astronomy Department, University of Washington, Box 351580, Seattle, WA 98195\\
$^2$Fermi National Accelerator Laboratory, MS 127, P.O. Box 500, Batavia, IL 60510}

\maketitle
\begin{abstract}
We present a combined analysis of all publicly available, visible HST observations
of transits of the planet HD 209458b.  We derive the times of transit,
planet radius, inclination, period, and ephemeris.  The transit times are then used to constrain the existence of secondary planets in the 
system.  We show that planets near an Earth mass can be ruled out in low-order
mean-motion resonance, while planets less than an Earth mass are ruled
out in interior, 2:1 resonance.  We also present a combined analysis of the transit times and 68 high precision radial velocity measurements of the system.  These results are compared to theoretical predictions for the constraints that can be placed on secondary planets.
\end{abstract}

\section{Introduction}

The discovery of 51 Peg B \citep{may95} fulfilled a many-decade quest to discover 
extrasolar planets around main sequence stars.  
The discovery of the first transiting planet HD 209458b \citep{cha00,hen00} allowed
the confirmation that radial velocity variations were indeed due to planets and
not some other period stellar phenomenon.  Currently ten
transiting planets have been found, allowing a measurement of the
mass-radius relation for short period extrasolar planets \citep{cha06}.
Besides confirming that radial
velocity variations are due to real objects with the sizes and masses of planets, 
transiting planets have allowed a host of other questions to be addressed, for
example, the planet atmospheric composition \citep{cha02}, constraints on the presence 
of planetary satellites \citep{bro01}, detection of thermal emission from the
planets \citep{dem05,cha05}, and the size of the planet core as a function 
of stellar metallicity \citep{gui06}.

In this paper we provide another application of transiting planetary systems:  we place a
constraint on the presence of terrestrial-mass planet companions in mean-motion
resonance with HD 209458b.  As shown in \citet{ago05}, a planet in resonance
induces a libration of the transiting planet with an amplitude 
$\sim 0.2 j^{-1} (M_2/M_1)
P_1$, where $M_1, P_1$ are the mass and period of the known transiting planet and $M_2$
is the mass of the companion planet in a $j$:$j+1$ resonance.  For HD 209458b, this
translates into a variation in the times of transit with a 3 minute amplitude and
few month period for $M_2=m_\oplus$.  Since this technique relies on a precise
measurement of the times of transit, we use the highest precision photometry available
for thirteen transits measured by the Hubble Space Telescope \citep{bro01,sch04,knu06}.
We carry out an analysis of the transit times with a careful quantifying of the
timing errors (\S \ref{dataanalysis}).  We use these results to constrain the
transiting planet parameters (\S \ref{results}).  We then use these derived times of transit
to place a constraint on a secondary coplanar planet (\S \ref{secondary}).
We finish up with a comparison of these results with those derived for the
TrES-1 transiting planet system (\S \ref{tres1}) and a comparison with the constraints
from radial-velocity measurements (\S \ref{rv}).

\section{Data Analysis} \label{dataanalysis}

At least ten science observing programs with the Hubble Space Telescope
have proposed to observe transits of the planet in the HD 209458 system.
The five ultraviolet observations do not have enough photons to measure 
the transit lightcurve precisely, while four programs have made 
observations of the planetary transit in the optical, of 
which three are now publicly available: 8789, 9171 and 9447 (the other program,
10145, has not been fully made public yet).  Two programs, Brown 8789 and Charbonneau
9447, utilized the STIS spectrograph as a CCD 
photometer.  By spreading the light from the star with a grism, in an 
effort to capture as many photons as possible, their relative photometry is likely the most 
precise ever obtained \citep{bro01,knu06}. 
The third program utilized the Fine Guidance Sensors to carry out rapid high
precision photometry \citep{sch04}. 
These programs give a total of thirteen transits observed with HST.  The primary
motivation for these observations was to constrain the 
planetary parameters, to detect absorption features in the planet's atmosphere, and
to search for satellites of the planet.  However, these data are 
also useful for timing each transit to measure a precise ephemeris \citep{wit05}
and to look for deviations from exact periodicity.  A summary of the observations is 
given in table \ref{tab01}.

\begin{table} 
\caption{Overview of the \textit{HST} observations of the HD 209458 system including the instrument, wavelength range, cadence, and date of observation.}\label{tab01}
\begin{tabular}{@{}lcccccll}
Transit & 
   Transit number &
    Orbits & 
       Instrument & 
              Filter/Grating & 
                      wavelength range & 
                                        exposure/readout time & 
                                                    date \cr
\hline
1 &  0& 5& STIS & G750M & 581.3-638.2 nm &  60/20 sec & UT 2000 April 25 \cr
2 &  1& 5& STIS & G750M & 581.3-638.2 nm &  60/20 sec & UT 2000 April 28-29 \cr
3 &  3& 5& STIS & G750M & 581.3-638.2 nm &  60/20 sec & UT 2000 May 5-6 \cr
4 &  5& 5& STIS & G750M & 581.3-638.2 nm &  60/20 sec & UT 2000 May 12-13 \cr
5 &117& 3& FGS  & F550W & 510-587.5 nm   &0.025/0 sec & UT 2001 Jun 11 \cr
6 &143& 2& FGS  & F550W & 510-587.5 nm   &0.025/0 sec & UT 2001 Sep 11 \cr
7 &160& 2& FGS  & F550W & 510-587.5 nm   &0.025/0 sec & UT 2001 Nov 10 \cr
8 &179& 2& FGS  & F550W & 510-587.5 nm   &0.025/0 sec & UT 2002 Jan 16 \cr
9 &252& 2& FGS  & F550W & 510-587.5 nm   &0.025/0 sec & UT 2002 Sep 30 \cr
10&313& 5& STIS & G430L & 290-570   nm   &  22/20 sec & UT 2003 May 3 \cr
11&321& 5& STIS & G750L & 524-1027  nm   &  19/20 sec & UT 2003 May 31 \cr
12&328& 5& STIS & G430L & 290-570   nm   &  22/20 sec & UT 2003 Jun 25 \cr
13&331& 5& STIS & G750L & 524-1027  nm   &  19/20 sec & UT 2003 Jul 5-6 \cr
\end{tabular}
\end{table}

\subsection{Reduction of pipeline data}

We re-reduced these data from the pipeline 
calibrated products that are available in the archive.   We followed
the reduction procedures described in \citet{bro01} 
and \citet{sch04}, 
but summarize the most important steps for each data set here.
For the STIS data we carried out the minimum reduction necessary to obtain
precise photometry.  We subtracted cosmic rays using
a 5 sigma rejection of the time series for each pixel within an
HST orbit, we binned the photons within 8 pixels of the spectrum 
peak (17 pixels wide) to derive the total counts for each frame, 
and assigned the midpoint of each exposure as the time for each frame. 
We discarded the first frame from each orbit.  For the first 
transit observed in program 8789 we only used the data for which the 
spectrum centroid was greater than 4 pixels from the edge of the CCD 
\citep[see discussion in][]{bro01} 
and we did not
utilize the first and last columns of the CCD which show larger
errors.  The 1-sigma errors on the flux are taken from the pipeline
calibrated errors which are summed in quadrature.
For the FGS data we took the sum of the four photometers, discarded
data which deviated more than 3 sigma from a fourth order fit
to the time series for each orbit, and binned the
data to 80 second bins (10 second bins gave the same results)
with the time stamp at the center of each bin.
With these time series we then fit a model for the transiting
planet and the flux sensitivity variations of Hubble, which we
describe next.

\subsection{Photometric Error bars}

The errors computed from the pipeline include counting noise and,
for the CCD data, read noise.  We examined the scatter in the data 
compared to the computed error bars and found that the scatter was
larger than the pipeline errors for the FGS data and second set of STIS data.
This could not be attributed solely to cosmic rays and we were not
able to identify the additional error source. Consequently, we estimated the errors 
for each transit individually by analyzing the out-of-transit data.

To account for the fluctuations in sensitivity of HST when
we determine the size of the error bars, we fit a model to the 
out-of-transit data for each transit with two components:  i) an
$n$th order polynomial fit as a function of the orbital phase of HST
ii) a linear fit as a function of time.   We measured the standard
deviation of the residuals of this fit in units of the statistical 
error of each point as a function of $n$, and measured the median 
of this standard deviation for all thirteen transits as a function 
of $n$.   We found that between $n=5$ and $n=6$ the median of the
standard deviation ratio was constant, indicating that including a 6th order 
term did not significantly improve the fit to the fluctuations in sensitivity 
as a function of the HST orbital phase.  The residuals did not correlate with
the HST orbital phase, so we then assigned the greater of the standard
deviation of the residuals for $n=5$ or the statistical error.  The 
first five transits (Table \ref{tab01}) had statistical errors nearly equal 
to the residuals,
while the error bars of the remaining transits were increased by 20-50\%.
These changes to the error bars were used throughout the remainder
of the analysis.

\subsection{Planet orbit and transit model}

We model the lightcurves by treating the planet and star as
perfectly spherical, and by assuming that the planet is on a circular 
orbit \citep{lau05,dem05}. 
Thus, there are only three free parameters to describe the
planet's orbit:  the ratio of the planet
to star radius, $R_p/R_s$;  the ratio of the relative
velocity of the planet and star to the stellar radius, 
$v/R_s$; and the inclination of the planet's orbit, $i$.
In addition, we allowed the time of each transit to
be a free parameter, giving thirteen transit times. From
these is derived the period of the planet, $P$, which then 
gives the ratio of the semi-major axis to the stellar radius,
$a/R_s=(v/R_s)P/(2\pi)$, necessary for computing the relative 
position of the planet and star.  In this approach the
actual values of the planet and stellar radii cannot
be determined, only their relative values \citep[to determine
the absolute values requires fitting the radial velocities
and using a stellar mass-radius relation as in][]{cod02,knu06}.

\subsection{Limb darkening}

If the lightcurves for each transit were only affected by
Poisson noise and were continuously sampled, then assumptions
about the shape of the transit lightcurve would be unnecessary:  
the mid-point of transit could be found by simply assuming the
transit lightcurve was symmetric about mid-transit.  However, the HST data are
affected by sensitivity variations due to temperature changes
in the telescope and by observing gaps, both due to Earth 
occultation.  This fact forces us to choose a model for the 
transit lightcurve in order to solve for the transit time, i.e. time
at transit midpoint.  The primary uncertainty in the modeling is the
shape of the stellar limb darkening.  To test the robustness
of the transit times to assumptions about limb-darkening,
we modeled the limb-darkening of the star with two approaches: 
stellar atmosphere predictions and quadratic limb-darkening.

\begin{figure}
\centerline{\psfig{file=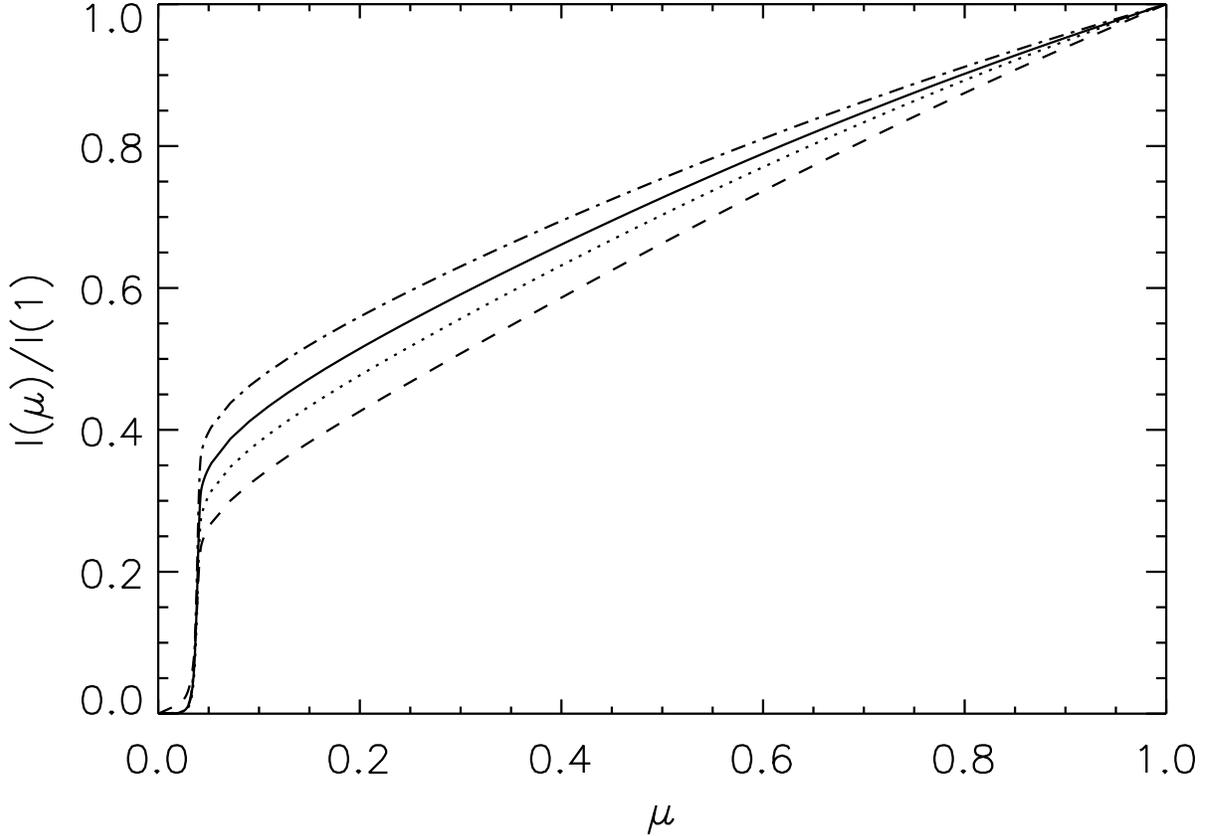,width=\hsize}} 
\caption{Limb darkening for the four instrumental configurations -
the vertical axis is the surface brightness of the star, $I$, as a 
function of $\mu=\cos{\theta}$, where $\theta$ is the polar angle
measured from the line of sight to the star.
Solid line is for STIS G750M, dotted line is for FGS F550W,
dashed line is for STIS G430L, while dash-dot line is for
STIS 750L.} \label{fig01}
\end{figure}

In the first case we used the computed limb-darkening from Phoenix NextGen
stellar atmosphere models \citep{cla03}. 
These models include an accurate, high-resolution treatment of 
the brightness at the limb using spherically-symmetric stellar models.
We used an atmosphere of 6100 K and a surface
gravity of $log(g)=4.5$, appropriate for HD 209458a 
\citep{val05}. 
We convolved the 
wavelength-dependent limb-darkening with the number of photons 
detected at each wavelength (for the FGS we convolved the FGS 
sensitivity with a stellar atmosphere to estimate the relative number 
of photons detected as a function of wavelength).  
The resulting limb darkening is shown in Figure \ref{fig01}.
There is only a gradual change in the limb darkening between
the four different instrument/filter combinations, but given
the timing precision required, these differences are significant.
We then used a cubic spline interpolation of the limb darkening 
to integrate over the portion of the stellar disk occulted by the planet.
The result is a more gradual ingress and egress than that predicted by 
analytic limb-darkening models which have a sharp edge.  

In the second case we assumed that the limb-darkening was a 
quadratic function in the cosine of the polar angle measured from
the line of sight.
We utilized the routines from \citet{man02} 
for computation of the lightcurves (this led to eight free
parameters for the four filter/instruments combinations used 
in the data).  We primarily used this case to check
that the timing results are insensitive to the accuracy
of the limb-darkening predicted from the atmosphere model,
discussed more below.

\subsection{HST flux sensitivity variations}

Thermal fluctuations during the orbit of HST lead to changes in
the sensitivity of each instrument at the 0.5\% level.  Fortunately
these variations appear to be fairly reproducible during
successive orbits (in addition to a smaller linear drift
between successive orbits), but they are not reproducible on longer 
timescales.  Following \citet{bro01} 
and \citet{sch04} 
we phased the observations of each transit to the HST orbital period 
($p_h$).  We used two models for the variation of the flux
sensitivity with orbital phase and two models for the ``secular" variation
with time over each transit, combined together giving a total
of four models for the background variations.  The orbital phase 
variation we modeled as i) a fifth-order polynomial, $\sum_{i=0}^4 c_i 
\phi^{(i+1)}$, where $0\le\phi<2\pi$ is the orbital phase of Hubble, or
ii) as a linear and two harmonic functions of orbital phase, 
$c_0\phi+c_1\sin(\phi)+c_2\cos(\phi) +c_3\sin(2\phi)+c_4\cos(2\phi)$.
The secular variation with time we modeled either as a linear function,
$c_5 f_{tran}(t)+c_6 t$, where $f_{tran}$ is the transit lightcurve
model normalized to unity outside transit, or a constant coefficient 
for each HST orbit, $c_{i+5} f_{tran,i}(t)$ where $i$ labels each 
HST orbit.  
The lightcurve model depends on sixteen non-linear parameters (the 3 
planetary orbital parameters and 13 transit times).  For the
models of the HST flux sensitivity variations there are either 
$7\times 13=91$ or $5\times 13+51=116$ linear parameters.
Coincidentally, the physically interesting parameters are non-linear,
while the uninteresting parameters are linear.

To most efficiently find
the best-fit for all 107 or 132 parameters, we used the following procedure:

1)  We carried out a minimization over the 16 non-linear parameters using
the MINPACK non-linear least squares routine \citep{cow84}. 

2)  The model lightcurve for each transit we computed by carrying
out a linear least-squares fit for the 91 or 116 linear parameters.

\noindent This procedure converged rapidly to the best $\chi^2$.  We mapped the
$\chi^2$ as a function of each non-linear parameter (while marginalizing
over the other parameters), finding that the $\chi^2$ has a quadratic
shape near the minimum, indicating that there is a unique solution.

\subsection{Timing Error computation}

The best-fit models have reduced chi-square in the range 
$\bar\chi^2=1.07-1.25$ for the four different flux sensitivity models.  
The reduced chi-squares
for the FGS data (which contain fewer data points) were as high as 2.6
(for the 5th transit) indicating that the models were a poor fit to
the data.  We examined the residuals of the FGS data and found that some 
residuals within the transits were correlated.  Thus
our best fitting models were not a sufficiently good description of
the data (either due to errors in the limb darkening or inaccuracies of the
modeling of the HST flux sensitivity variations).   Therefore, to obtain a 
better estimate of the error bars, taking into account these systematic 
effects, we used a Monte Carlo bootstrap simulation of the errors in which we 
shifted the residuals by a random number of points for each transit so that 
the correlations were maintained.  We also reversed the residuals for each 
orbit and repeated the shift.  We followed this procedure 200 times, added the 
shifted residuals to the best-fit model, and then re-fit using the same procedure 
that was applied to the original data described in the previous section.  Our 
expectation is that this procedure gives an estimate of the errors due to 
inaccuracies in our assumed flux sensitivity models.  We found that the resulting 
errors on non-linear model parameters, such as transit times, were larger by a 
factor of a few than if 
we simply used Poisson errors for the bootstrap or resampled the residuals randomly.

Although this process gives an indication of the size of the systematic errors,
we also assessed the systematic errors by comparing the results from the
four different HST flux sensitivity variation models.  The differences 
in $\chi^2$ between the four different models was small -
all had reduced $\chi^2$ near unity, so it is difficult to favor one
particular sensitivity variation model.  However, the differences between
the derived transit times for each model was larger than the size of the 
error bars for a given model.  We believe that this discrepancy is
probably due to imperfections in all of the four flux sensitivity models,
so we have taken the mean and standard deviations of all eight hundred
Monte-Carlo simulations as an estimate of the transit times and their
errors.  Finally, to check the robustness of the stellar atmosphere
limb-darkening we have used the quadratic limb-darkening law with
the polynomial function of phase and linear secular term to fit
the transit times.  We find that these times agree within the errors 
with the times derived using the stellar atmosphere model and the same flux 
sensitivity model, so we conclude that the stellar atmosphere limb-darkening model
provides an accurate estimate of the transit times of the data.

To these measured transit times we apply a correction for the
motion of the Earth about the barycenter of the solar system
(a heliocentric correction is insufficient as it differs from
barycentric by up to 5 seconds).  The motion of the Hubble Space Telescope 
about the Earth contributes a negligible timing error.
The resulting transit times are reported in Table \ref{tab02}.
We find from the Monte Carlo simulations that there is no significant
correlation between the individual transit times so we report the 
standard deviation of each time without the full covariance matrix.
Our fitted transit times agree well with those obtained by \citep{knu06}, 
most being within 1/2 $\sigma$.  The largest discrepancies are for 
transits 12 and 13 which disagree by 1.25 $\sigma$ and 0.75 $\sigma$ 
respectively.  In the analysis by \cite{knu06} it is these same transits 
which deviate most significantly from a constant period.

\begin{table}
\caption{Transit time and uncertainty in days and in seconds for the 13 observed transits.}\label{tab02}
\begin{tabular}{@{}lccc}
Transit & Transit time    & Error & Error     \cr
        & (- 2450000 BJD) & (days)&(seconds) \cr
  1 &    1659.93678 &    0.00015 &  12 \cr
  2 &    1663.46150 &    0.00025 &  21 \cr
  3 &    1670.51102 &    0.00013 &  11 \cr
  4 &    1677.56044 &    0.00037 &  31 \cr
  5 &    2072.33284 &    0.00049 &  42 \cr
  6 &    2163.97574 &    0.00028 &  24 \cr
  7 &    2223.89685 &    0.00058 &  50 \cr
  8 &    2290.86741 &    0.00051 &  44 \cr
  9 &    2548.17311 &    0.00029 &  24 \cr
 10 &    2763.18306 &    0.00018 &  15 \cr
 11 &    2791.38087 &    0.00013 &  10 \cr
 12 &    2816.05431 &    0.00021 &  18 \cr
 13 &    2826.62867 &    0.00017 &  14 \cr
\end{tabular}
\end{table}

\begin{figure}\label{fig02}
\centerline{\psfig{file=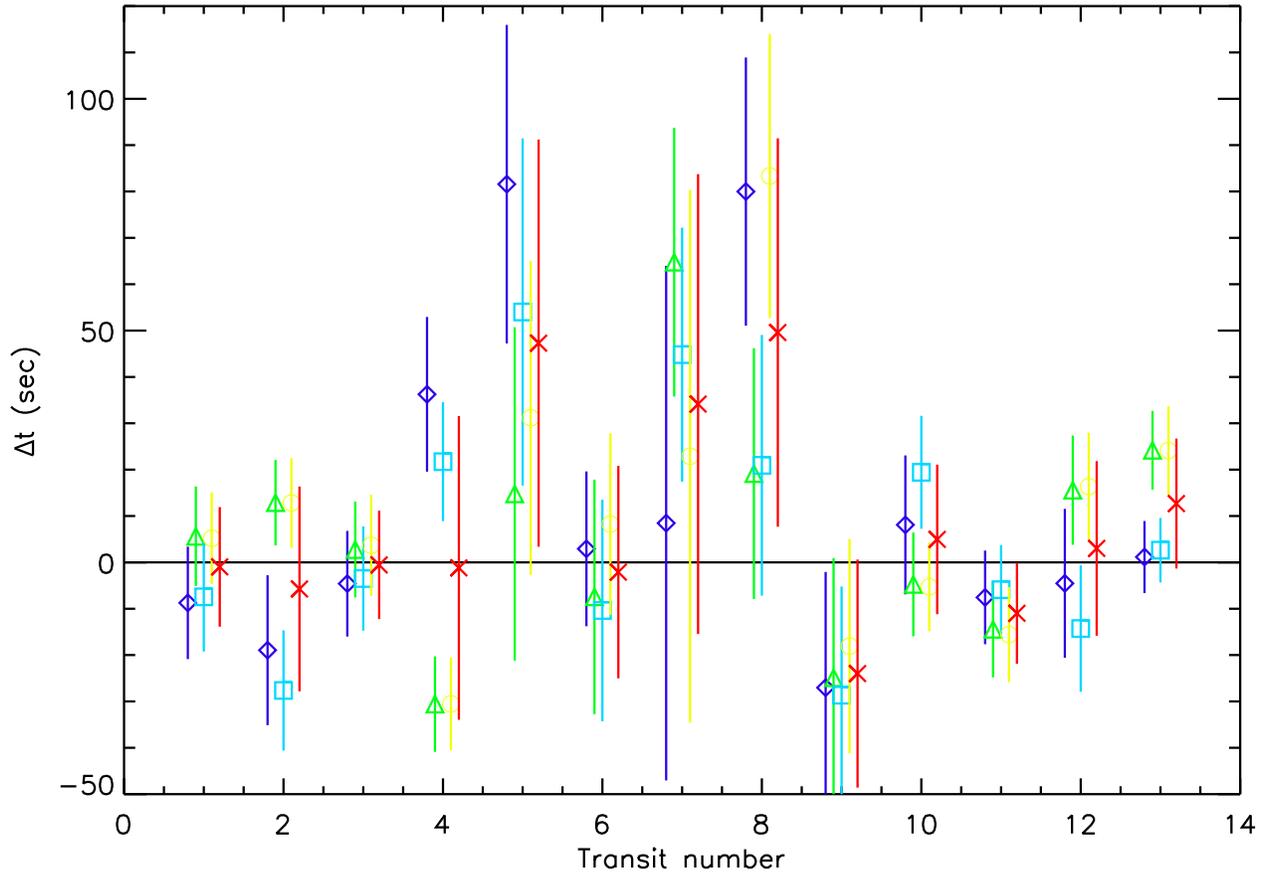,width=\hsize}} 
\caption{Differences of transit times from uniform period versus
number of transit (each point is shifted horizontally slightly for 
clarity).  Green/triangle- polynomial
and linear; yellow/circle - harmonic and linear; dark blue/diamond - polynomial and
flux offset for each orbit; light blue/square - harmonic and flux offset for 
each orbit.  Red/cross points are mean and standard deviation of all four 
flux sensitivity models.}
\end{figure}

\section{Results} \label{results}

\begin{figure}\label{fig03}
\centerline{\psfig{file=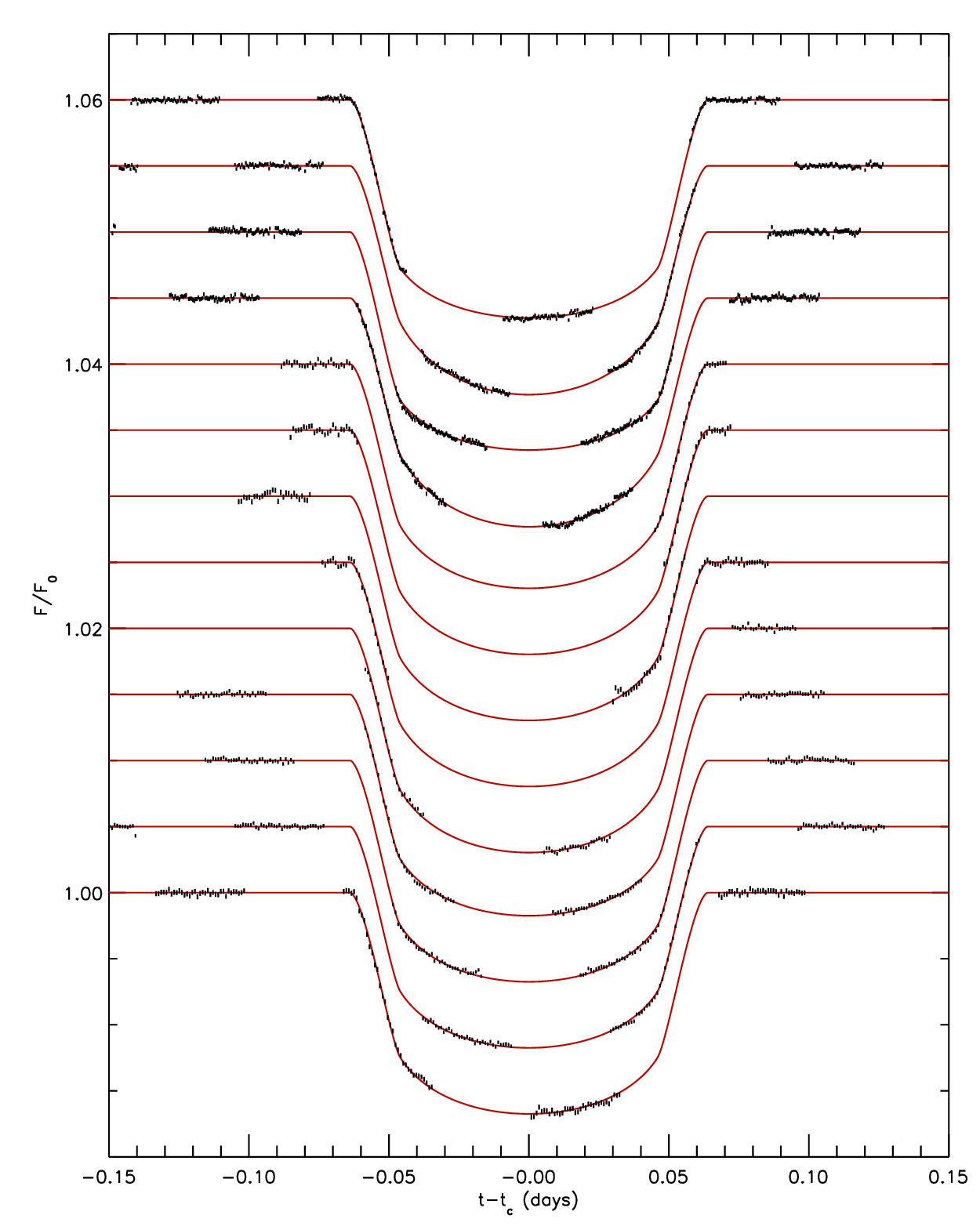,width=\hsize}} 
\caption{Best-fit models overplotted with data (some data -0.15 days
before mid-transit is
off the plot).  The data points are plotted with error bars, while the
best-fit model is plotted as a red line.  Each transit has been normalized
to one out of transit and shifted by 0.05;  the bottom is transit 1 and
the top is transit 13.}
\end{figure}

The results of the best fit models are shown in Figure \ref{fig03}.

\subsection{Constraints on radius, inclination}

\begin{figure}
\centerline{\psfig{file=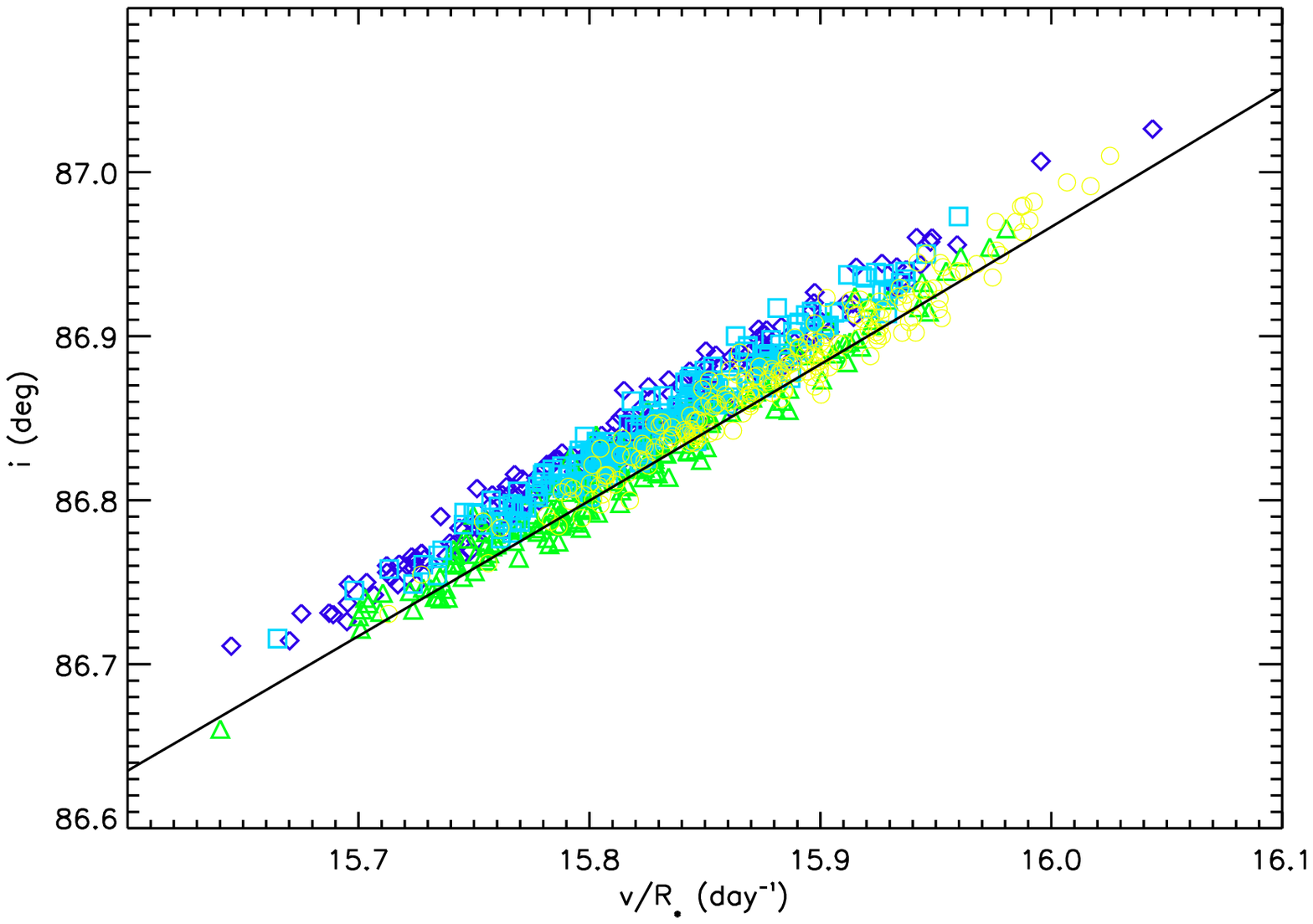,width=\hsize}} 
\caption{Velocity versus inclination - Monte Carlo (green/triangles - polynomial
and linear; yellow/circles - harmonic and linear; dark blue/diamonds - polynomial and
flux offset for each HST orbit; light blue/squares - harmonic and flux offset
for each HST orbit) and analytic expectation (black line).}\label{fig04}
\end{figure}

Since the data consist of a time series, the best constrained parameters
are the period, $P$, and the duration of each transit, $T_t$.  This leads to 
a degeneracy between the physical parameters: inclination
and velocity.  The relation between velocity and inclination is given
by
\begin{equation}\label{degeneracy}
v = R_* \left({T_t^2 \over 4} +{P^2 \over 4\pi^2} \cos^2{i}\right)^{-1/2}.
\end{equation}
As long as $v$ and $i$ obey this relation, $P$ and $T_t$ are unchanged.
Thus, for a very small uncertainty in $T_t$ and $P$, there can be a
larger uncertainty in $i$ and $v/R_*$.
The inclination is constrained by the shape of the limb-darkening
and the duration of ingress and egress, which partially (but not completely)
break the velocity-inclination degeneracy.
In Figure \ref{fig04} we show the results of our fitting for $\hat v = v/R_*$
and $i$.   We plot the results of Monte Carlo simulations of each
of the four models discussed above.  The correlation between
velocity and inclination follows equation \ref{degeneracy} as expected.

We derive a radius of $R_p/R_*=0.1201\pm 0.0006$, consistent at
1$\sigma$ with the radius derived in \citet{man02} (our derived
error is larger since we have taken into account correlated photometric errors
using the Monte Carlo analysis). 
We derive a velocity of $v/R_* = 15.835\pm 0.065$ days$^{-1}$,
which combined with the period gives the ratio of the size
of the semi-major axis to the size of the star, $a/R_*=8.883\pm.036$.
We derive an inclination of $i=1.5157\pm0.00095$
or $i=86.845^\circ\pm0.055^\circ$ which is consistent with that
reported in \citet{bro01}, although our derived error bar is
smaller due to our using a fixed limb-darkening law.  Since one
can only derive dimensionless parameters or parameters with 
dimensions of time or flux directly from the lightcurve, we cannot
measure the absolute size of the star, orbit, or planet.

\subsection{Ephemeris}

The accurate and realistic estimation of errors on the transit times
allows for an accurate estimate of the ephemeris of the transiting
planet.  We find values of:  $T_0=2451659.93677 \pm  0.00009$ BJD
(error is 8 seconds)
and $P=3.5247484  \pm 0.0000004$ days  (error is 0.03 seconds).
Our derived ephemeris is in good agreement with the ephemeris
derived by \cite{knu06}, differing by less than 1/2 $\sigma$ for both 
$P$ and $T_C$.  We note, however, that the small uncertainty in the 
period should be considered the uncertainty in the mean period that spans 
the modest time scale over which the data were obtained.  Current constraints 
on secondary planets in this system cannot rule out bodies that could change 
either the mean period or the instantaneous period of HD209458b by more than 
the quoted error.  These changes could happen over time scales longer than 
the span of the observations, such as a very small companion in a mean-motion 
resonance, or over shorter time scales, such as a companion on a large, highly 
eccentric orbit.

The period and zero-point of the ephemeris are anticorrelated with a 
correlation coefficient of -0.8;  this is due to the fact that an
increased zero point requires the period to shrink to go through
the later data points.  The anti-correlation decreases to -0.2 if
the zeroth eclipse is chosen to be halfway between the first and last
eclipses.  Given the estimated statistical plus
systematic errors, the chi-square is $\chi^2 = 6.0$ for eleven
degrees of freedom (thirteen transits minus two fitting parameters).
This indicates that the transit times are consistent with being
uniform and that there may not be a second planet
detectable with the current HST data, as concluded by \citet{knu06}.  
In the next section we use these data to constrain the masses of 
additional planets in the system.

\section{Constraints on Secondary Planets} \label{secondary}

\begin{figure}
\centerline{\psfig{file=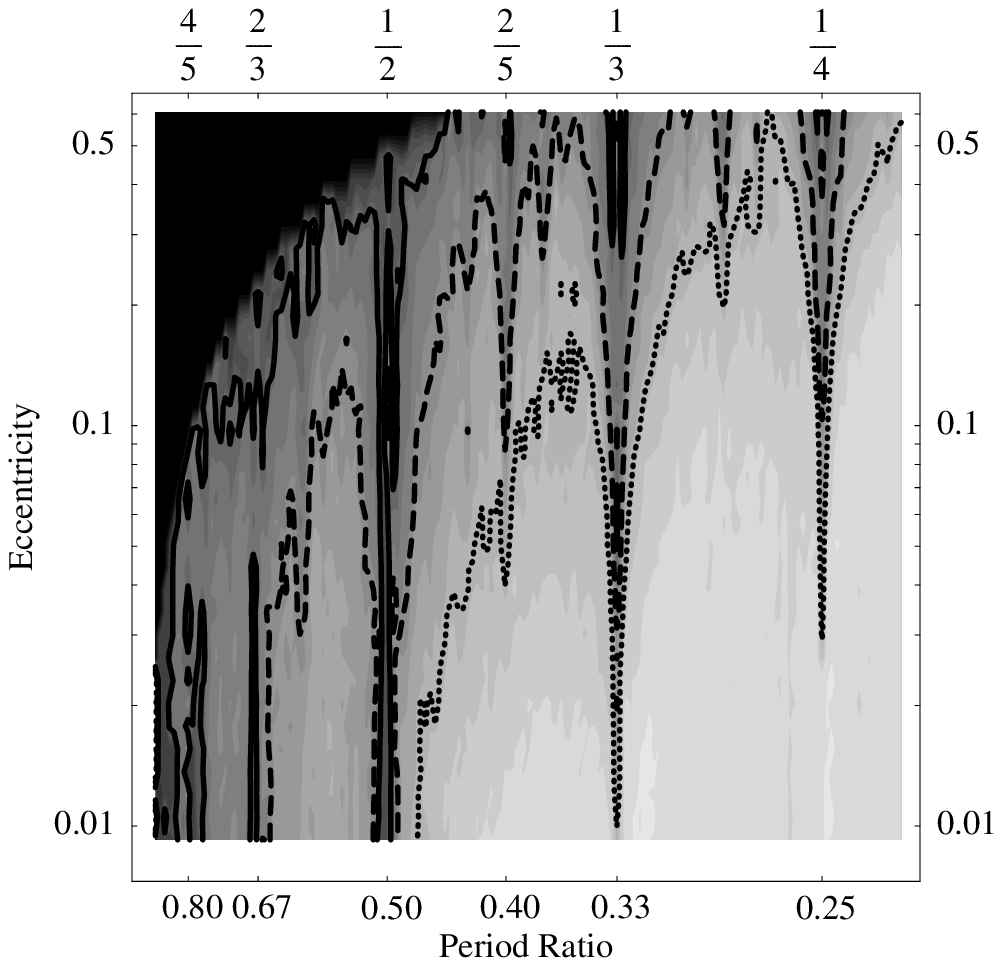,width=\hsize}} 
\caption{Constraints on the mass of an interior, secondary planet in the HD209458 
system as a function of semi-major axis ratio and eccentricity of the secondary 
planet.  We assume that the known planet has an initial eccentricity of zero.  
The contours correspond to 100 (dotted), 10 (dashed), and 1 (solid) earth-masses.  
The dark region in the upper-left portion of the graph is where the orbits overlap.} \label{fig05a}
\end{figure}

\begin{figure}
\centerline{\psfig{file=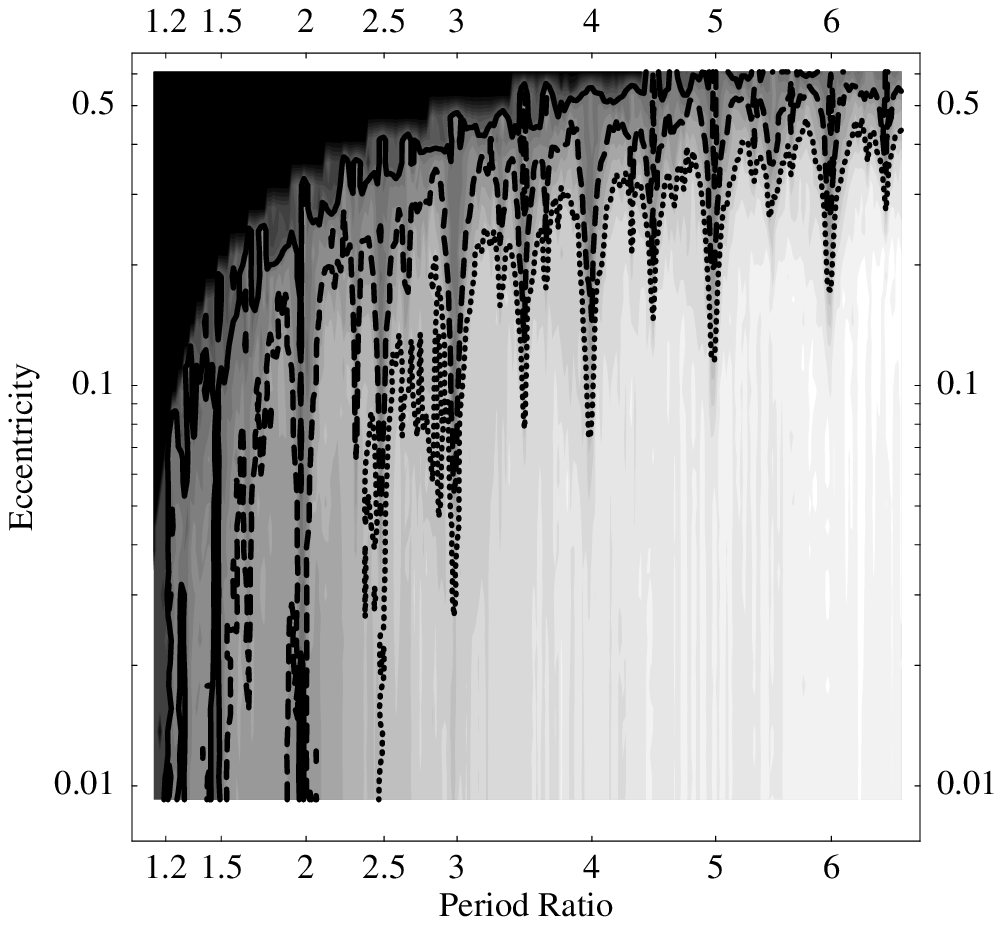,width=\hsize}} 
\caption{Constraints on the mass of an exterior, secondary planet in the HD209458 
system as a function of semi-major axis ratio and eccentricity of the secondary 
planet.  We assume that the known planet has an initial eccentricity of zero.  
The contours correspond to 100 (dotted), 10 (dashed), and 1 (solid) earth-masses.  
The dark region in the upper-left portion of the graph is where the orbits overlap.} \label{fig05b}
\end{figure}

Following the procedure outlined in \cite{ste05} we use the measured transit times
to constrain the presence of secondary planets in the HD209458 system.  In brief, 
for a given value of the semimajor axis and eccentricity of a putative second 
coplanar planet we determine the maximum mass that a companion could have without causing 
a significant deviation (3-$\sigma$) from the data.  Figures \ref{fig05a} and 
\ref{fig05b} show the maximum mass that a perturbing planet can have in this system 
for an interior perturbing planet and an exterior perturbing planet respectively.

We see from these figures that near several mean-motion resonances, a secondary planet 
must be comparable to or less than the mass of the Earth for various values of the 
eccentricity.  This is particularly manifest in the 2:1 and 3:2 mean motion resonances 
where this technique is most sensitive.  Here, a secondary planet can be no larger than 
$5M_{\oplus}$ for an exterior perturber, $2M_{\oplus}$ for an interior perturber in the 
3:2 resonance, and $< 1M_{\oplus}$ for an interior perturber in the 2:1 resonance.  These 
constraints are independent of the eccentricity of the perturber.  For low initial
eccentricity orbits ($e\sim 0.01$) the limits are as low as $0.2M_{\oplus}$, twice the 
mass of Mars.  

\begin{figure}
\centerline{\psfig{file=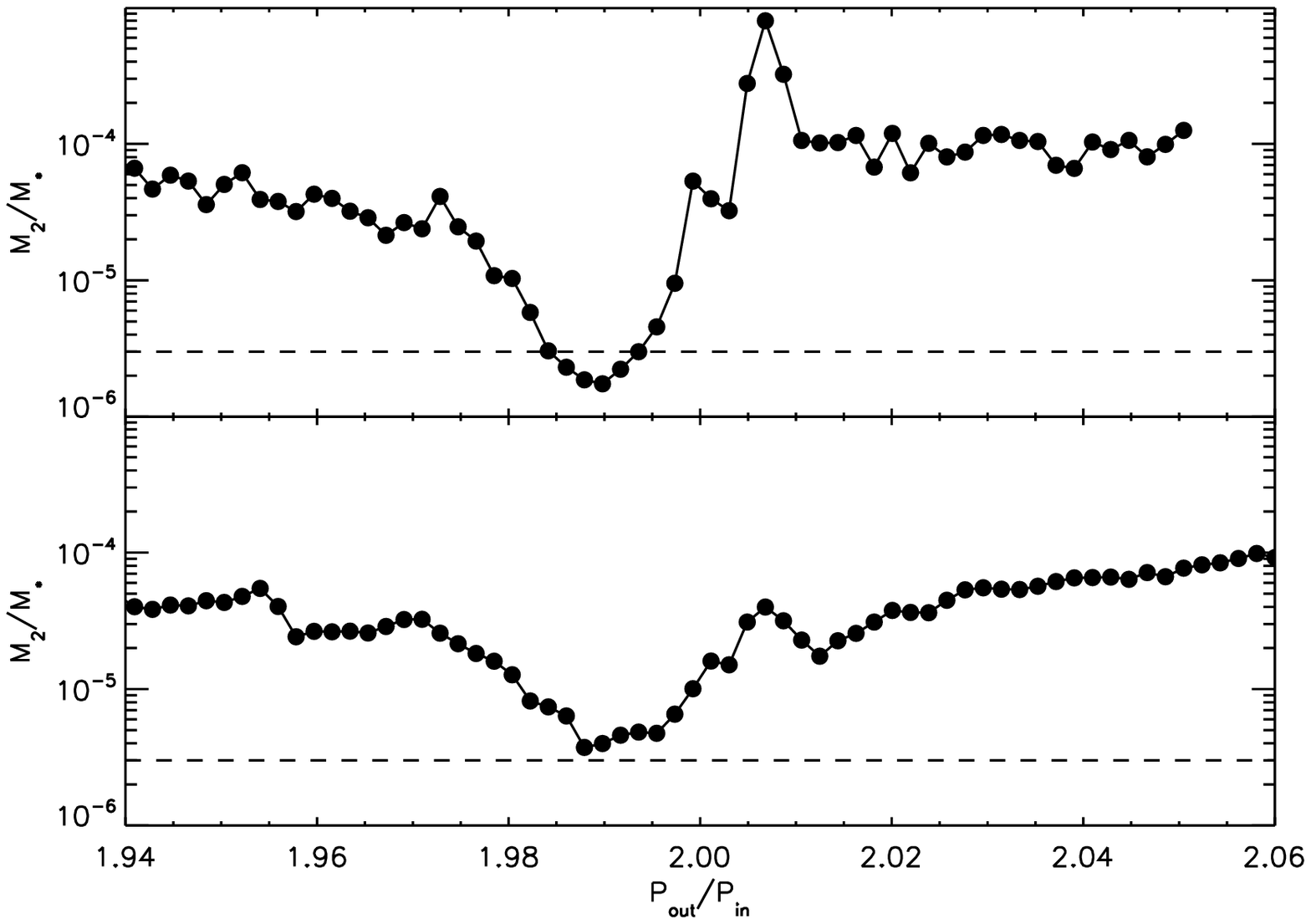,width=\hsize}}
\caption{Upper limit on the mass of any coplanar planetary companion present in the interior
resonance (upper panel) or exterior (lower panel) 2:1 resonance
with HD 209458b.  The horizontal dashed line corresponds to an earth-mass plant.}\label{fig06}
\end{figure}

We marginalized over eccentricity to compute limits on the presence of
coplanar planetary companions in a 2:1 and 1:2 resonance with the transiting planet.
Figure \ref{fig06} shows the limits on planets within these resonances.
Remarkably, the interior resonance can rule out at 3-$\sigma$ the presence of 
earth-mass bodies of any eccentricity within 0.5\% of exact resonance (we have not
included bodies with eccentricities large enough that their Hill spheres overlap 
with that of HD 209458b).  The exterior resonance is slightly less sensitive,
but can still rule out $\sim 2M_\oplus$ mass planets within 0.5\% of resonance.
We have checked for stability on timescales of several hundred orbits for 
resonant bodies, but have not carried out a longer timescale stability analysis.

\subsection{Comparison with TrES-1 System} \label{tres1}

The first analysis of this sort was conducted for the TrES-1 planetary system \citep{ste05}.
That analysis used 11 transits observed with ground-based telescopes and was 
also able to probe for sub-earth mass planets.  Those observations had an average timing 
uncertainty near 110 seconds--the most precise being 26 seconds.  For this analysis of 
HD 209458b, being observed by \textit{HST}, the average timing uncertainties are 
significantly smaller, with an average of 25 seconds (17 seconds for the STIS data).

\begin{figure}
\centerline{\psfig{file=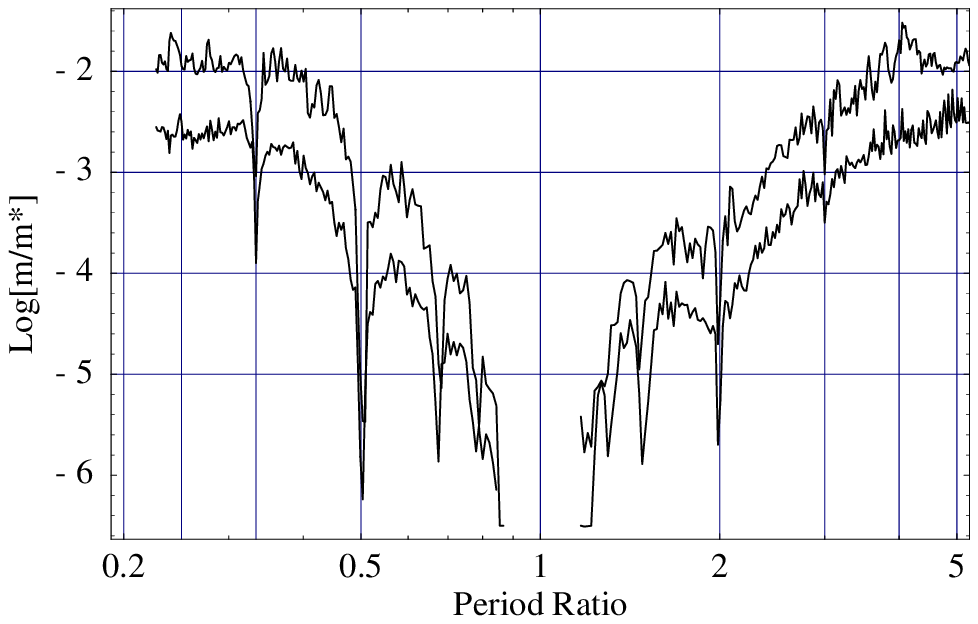,width=\hsize}} 
\caption{A comparison of the limits on a secondary planets obtained by TTV analysis for 
the TrES-1 system with 11 transits (upper curve) and the HD209458 system with 13 transits 
(lower curve).  The primary distinction between the transit data for each system is that 
the ground-based TrES-1 observations have a timing uncertainty that is a factor of four, 
on average, larger than those of the HD209458 system, which were observed with \textit{HST}.} \label{fig07}
\end{figure}

The factor of four improvement in timing precision corresponds to a comparable improvement 
over the constraints on secondary planets in the HD209458 system.  Figure \ref{fig07} 
compares the limit on the mass of a companion planet for the TrES-1 system with that of 
the HD209458 system for perturbers with an eccentricity of 0.02.  This figure shows that 
the constraint on the HD209458 system is similar in shape but markedly lower.

Though the space-based data can give tighter constraints for systems like TrES-1 and HD209458, 
figure \ref{fig07} demonstrates the capacity that ground-based telescopes have to probe for 
secondary planets.  The transit observations of the TrES-1 system were conducted on relatively 
modest telescopes (1m) and were not optimized for obtaining the time of transit.  Even then, 
the best of the ground based observations 
\cite[obtained using the 1.2m Fred Whipple telescope][]{cha05} have a timing precision that is 
essentially equal to the average uncertainty of 
the \textit{HST} data.  Thus, additional ground-based observations, with precision comparable 
to the best of those obtained for the TrES-1 system could readily
probe for companions with masses less than that of the Earth that are trapped in 
low-order, mean-motion resonances.

\subsection{Constraints Including Radial Velocity Measurements} \label{rv}

Theoretical studies \citep{ago05} indicate that high-precision radial velocity measurements should 
provide more stringent constraints than TTV for planets that are not in resonance.  We therefore 
conducted a combined analysis that utilizes both the transit data and radial velocity data for 
this system.  We used the radial velocity data that were published by \citet{lau05}; 
discarding the data that they indicate were taken near transits.

\begin{figure}
\centerline{\psfig{file=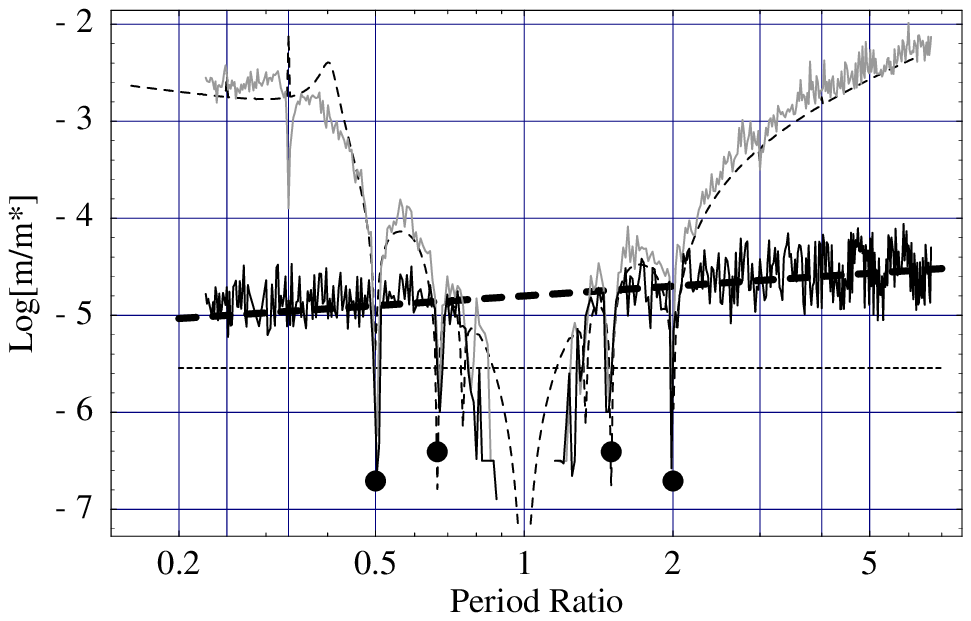,width=\hsize}} 
\caption{Limit on the mass of a low-eccentricity secondary planet in the HD209458 system.  The solid, black curve is 
obtained from an analysis of transit times and radial velocity measurements simultaneously assuming 
that the orbits are initially circular.  The gray curve is the result of an analysis of the transit 
times only, with the perturbing planet initially having an initial eccentricity of 0.02.  The thin dashed 
curve is from a perturbation theory calculation and are given by equations (A7) and (A8) in 
Agol et al. (2005). 
The large dots are from equation (33) in Agol et al. (2005), although without the
factor of 4.5 which is the difference between the peak TTV amplitude and the
TTV amplitude for intitially exact $j$:$j+1$ resonance.
The large, dashed line is from equation (2) in Steffen \& Agol (2005)
and is the expected limit from radial velocity measurements.  The small, horizontal dashed line corresponds to the mass of the Earth.} \label{fig08}
\end{figure}

For this combined analysis we applied essentially the same algorithm that we used for the 
transit-only analysis and for the analysis of the TrES-1 system in \cite{ste05} with the difference
that we compare the radial velocity data to a model with two Keplerian orbits with the 
parameters of the two planets.  The change in the orbital elements of the planets due to 
non-resonant planet-planet interactions are small enough to not affect the radial velocities, justifying
the use of Kepler's equation. 
The results of our combined analysis are shown in Figure \ref{fig08}.
We see from Figure \ref{fig08} that our results are in good agreement with the theoretical predictions 
given in \cite{ago05} and \cite{ste05}.

\section{Discussion}

This investigation started as part of a larger quest for detecting terrestrial-mass planets orbiting
main-sequence stars.  Transits times are particularly sensitive to earth-mass planets in
mean-motion resonance with the transiting planet.  For the case of the HD 209458 system, we can rule out a wide range of terrestrial-mass planets based on the \textit{HST} data analyzed
and presented here.   The paradigm of core accretion plus migration predicts that
smaller planets ought to be trapped in resonance with larger, short-period planets
\citep{zho05,tho05}.  With the analysis of only two systems (HD 209458 and TrES-1), we cannot draw general
conclusions, but in the next few years there will be likely be transit timing observations
of all currently known transiting planets, so a significant constraint on the
presence of resonant, earth-mass planets could be placed with this sample.
In the process we have derived accurate parameters for the
transiting planet which are broadly consistent with the analyses of other authors.

Despite its lofty position above the Earth's atmosphere, it is 
apparent from these data that the Hubble Space Telescope is not
ideally suited to timing transits.  The main problem is its
low-Earth orbit which causes occultation of the Sun leading to
thermal variations within the orbit and causes occultation of
the target star leading to gaps and re-pointing.  The thermal
variations cause sensitivity variations, while the gaps make
a model for limb-darkening necessary to measure the transit
times.  If the data errors behaved according the Poisson statistics,
the expected timing uncertainty
for the STIS data would be of order 
3-5 seconds, while the derived times are 4-5 times larger.  The
FGS data had a decreased sensitivity relative to the STIS data
due to a lower count rate and the observation of each transit
for only two or three orbits, an observing strategy which conserved
Hubble time but was unsuited to the precise measurement of transit 
times.  Future, precise measurements of transit times are possibly 
better carried out with satellites in Earth-trailing orbit, such
as the Spitzer Space Telescope.  Spitzer has the extra advantage
that the infrared is less subject to limb-darkening; however, it
would collect fewer photons as it has a smaller aperture and is
observing past the peak of the photon count spectrum of G-type
stars. In addition, ground-based observations with accurate relative
photometry and coverage from observatories separated in longitude
should make very accurate transit timing feasible from the ground.
Finally, the Kepler satellite will discover a large number of
transiting jupiters - the longer period planets will allow even
more precise constraints on companions over a larger range of
semi-major axis ratios \citep{hol05}.   To contrain the presence of planets with
mutual inclinations, more data points are required so that the
model can accomodate more free parameters.

\section{Acknowledgements}

JHS acknowledges support from NASA Graduate Student Research Fellowship
grant number NN605GO11H and from the Brinson Postdoctoral Fellowship.  
Support for this work was provided by NASA
through grant number HST-AR-10637.01-A from the Space Telescope Science
Institute which is operated by the Association of Universities for
Research in Astronomy, Inc., under NASA contract NAS5-26555.  We acknowledge
Alfred Schultz and Wayne Kinzel for providing us with their preliminary
reduction of the FGS data.  We also thank Peter Hauschildt for making numerical
limb-darkening models freely available. This paper has been assigned
 DOE preprint number FERMILAB-PUB-06-279-A-CD.

\bibliography{agol_steffen}
\clearpage

\end{document}